\documentclass[12pt]{iopart}

\usepackage{iopams}  
\usepackage{graphicx}
\usepackage{siunitx} 
\usepackage{hyperref}
\usepackage{cite}
\graphicspath{{Pictures/PNG/}{Pictures/PDF/}{Pictures/JPG/}{Pictures/}}

\begin{document}

\title{Creation of double-well potentials in a surface-electrode trap towards a nanofriction model emulator}

\author{U. Tanaka$^{1,2,3}$, M. Nakamura$^1$, K. Hayasaka$^{3,1}$, A. Bautista-Salvadora$^{4,5}$, C. Ospelkaus$^{4,5}$, and T. E. Mehlst\"aubler$^{4,5}$ }    
\address{$^1$Graduate School of Engineering Science, Osaka University, 1-3 Machikaneyama-cho, Toyonaka 560-8531, Japan}
\address{$^2$Center for Quantum Information and Quantum Biology, Institute for Open and Transdisciplinary Research Initiatives, Osaka University, Osaka 560-8531, Japan}
\address{$^3$National Institute of Information and Communications Technology, 588-2 Iwaoka, Iwaoka‑cho, Nishi‑ku, Kobe 651-2492, Japan}
\address{$^4$Physikalisch-Technische Bundesanstalt, Bundesallee 100, 38116 Braunschweig, Germany}
\address{$^5$Institute of Quantum Optics, Leibniz Universit\"at Hannover, Welfengarten 1, 30167 Hannover, Germany}
 
\ead{utako@ee.es.osaka-u.ac.jp}
\vspace{10pt}
\begin{indented}
\item[]October 2020
\end{indented}

\begin{abstract}
We demonstrate a microfabricated surface-electrode ion trap that is applicable as a nanofriction emulator and studies of many-body dynamics of interacting systems. The trap enables both single-well and double-well trapping potentials in the radial direction, where the distance between the two potential wells can be adjusted by the applied RF voltage. In the double-well configuration, parallel ion strings can be formed, which is a suitable system for the emulation of the Frenkel-Kontorova (FK) model. We derive the condition under which the trap functions as a FK model emulator. The trap is designed so that the Coulomb interaction between two ion strings becomes significant. We report on the microfabrication process for such downsized trap electrodes and experimental results of single-well and double-well operation with calcium ions. With the trap demonstrated in this work we can create atomically accessible, self-assembled Coulomb systems with a wide tuning range of the corrugation parameter in the FK model. This makes it a promising system for quantum simulations, but also for the study of nanofriction in one and higher dimensional systems.
\end{abstract}

%
\vspace{2pc}
\maketitle
%
%

\section{Introduction}

Trapped atomic ions are a well-isolated and controllable system that can be used to model a physical system of interest and to investigate the physics of large and computationally intractable systems. An interesting system that can be implemented using trapped ions is the Frenkel-Kontorova model, which describes the emergence of nanofriction. At atomic scales, a deviation from the Amontons-Coulomb law was theoretically predicted, and the atomistic models had been investigated with macro crystals such as a NaCl crystal ~\cite{Socoliuc04} and graphite flakes~\cite{Dienwiebel04}. However, pure atomic systems at the nanoscale are hard to access with macro crystals. Due to the development of the field of cold atoms, manipulation of small numbers of atomic ions, from single to hundreds of ions, became possible. Nanofriction model emulators using cold ions were proposed and numerically analyzed \cite{Mata07, Benassi11, Haffner11}, and several pioneering experimental studies have been reported ~\cite{Vuletic15,Vuletic16,Vuletic17,Mehlstaubler17,Schmirander17}. Harmonic chains in a corrugation potential were emulated by using an ion string in an optical lattice~\cite{Vuletic16} and a zig-zag ion configuration in a three-dimensional linear Paul trap~\cite{Mehlstaubler17, Schmirander17}. The latter system makes use of the Coulomb interaction between two rows of ions, therefore, it includes the effect of back-action from the other “surface”. These studies are of substantial significance because they bridge the gap between nanoscale and macroscale. 

Recent progress in microfabricated surface-electrode ion traps enables flexible ion configurations including two-dimensional ion arrays. There are two types of surface-electrode traps to realize a two-dimensional configuration: arrays of quadrupole Paul traps ~\cite{Leibfried09_2D, Blatt11_2D, Hensinger14, Schaetz16} and that of linear Paul traps~\cite{Wunderlich11, Tanaka14}. In the case of quadrupole Paul traps, methods for optimizing geometries have been developed by focusing on the maximization of the electric field curvature of individual trapping sites for arbitrary ion heights and separations ~\cite{Leibfried09_2D}. A 2D lattice consisting of Yb$^+$ ions ~\cite{Hensinger14} and that of Mg$^+$ ions ~\cite{Schaetz16} have been demonstrated using microfabricated ion traps. As for the geometry based on a linear surface-electrode trap, multiple parallel traps have been proposed and detailed spin-spin coupling strengths have been estimated ~\cite{Wunderlich11}. Demonstration of two parallel ion strings have been reported by us in ~\cite{Tanaka14}.
The advantages of our concept are as follows: i) the trap potential can be adjusted to be either single-well or double-well by varying the RF voltages, and ii) the separation between two ion crystals can be adjusted over a certain range by varying the RF voltages. The full control of independent 1D to 3D ion Coulomb crystals with tunable interaction strength between them can be useful for the study of nanofriction in different dimensions.

Here, we report on a novel ion trap produced with an improved fabrication process that allows down-sizing our design. With this micro-engineered surface-electrode trap coupling neighboring ion strings in a controlled way becomes possible. We propose to use this new platform for studies of nano-tribology in 1D to 3D systems using trapped ions.We present our trap design, describe the analysis of the trap potential, adjustable parameters and the required conditions to observe the coupling between two ion strings. The trap presented here is half the size of our previous trap~\cite{Tanaka14}. We describe the method and results of microfabrication and perform trapping experiments with calcium ions. We calculate the Coulomb potential generated by parallel ion strings in our trap and show that the system is applicable to a nanofriction emulator. Besides the nanofriction model, the surface-electrode trap reported here offers the possibility of scaling-up quantum simulations.

\subsection{Trap Design}
\begin{figure}
\centering
\includegraphics[width=1\columnwidth]{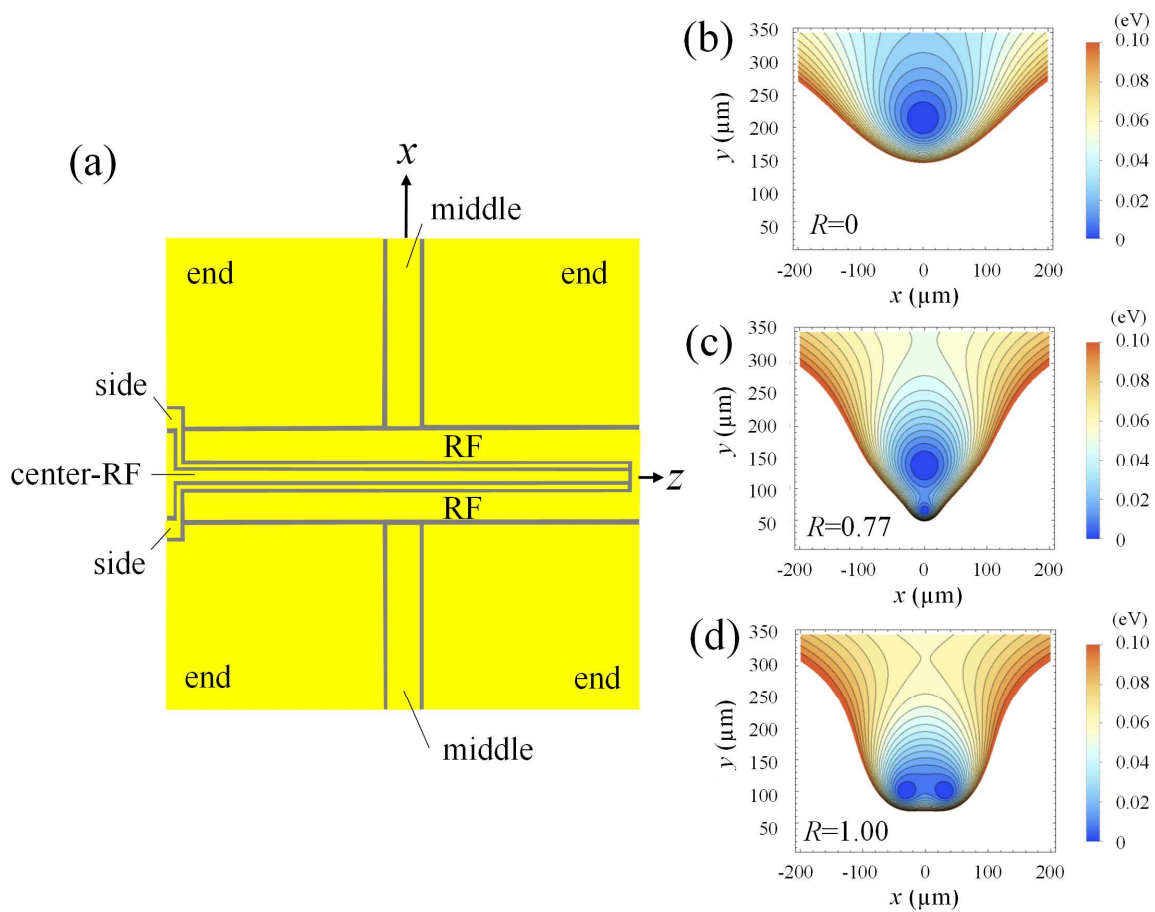}
\caption{(a) Schematic illustration of the trap layout. Not to scale. A RF voltage ($V_{RF}$) is applied to the two RF electrodes to provide radial confinement. In addition, a RF voltage is applied to the center-RF electrode ($V_{cRF}$) at a certain ratio $R$ which is equal to $V_{cRF}/V_{RF}$. Confinement along the z direction is achieved by applying DC voltages to end and middle electrodes. Side electrodes are used for adjusting DC potential to reduce the micromotion. (b)(c)(d) Calculated pseudopotential for different values of $R$. The dimension of electrodes and condition of applied RF voltages are described in the text. The plot is shown up to 0.1 eV. The white region near the trap surface indicates that the potential energy is larger than 0.1 eV. The separation between contours is 5 meV.}
\label{TrapLayout} 
\end{figure}

\begin{figure}
\centering
\includegraphics[width=0.5\columnwidth]{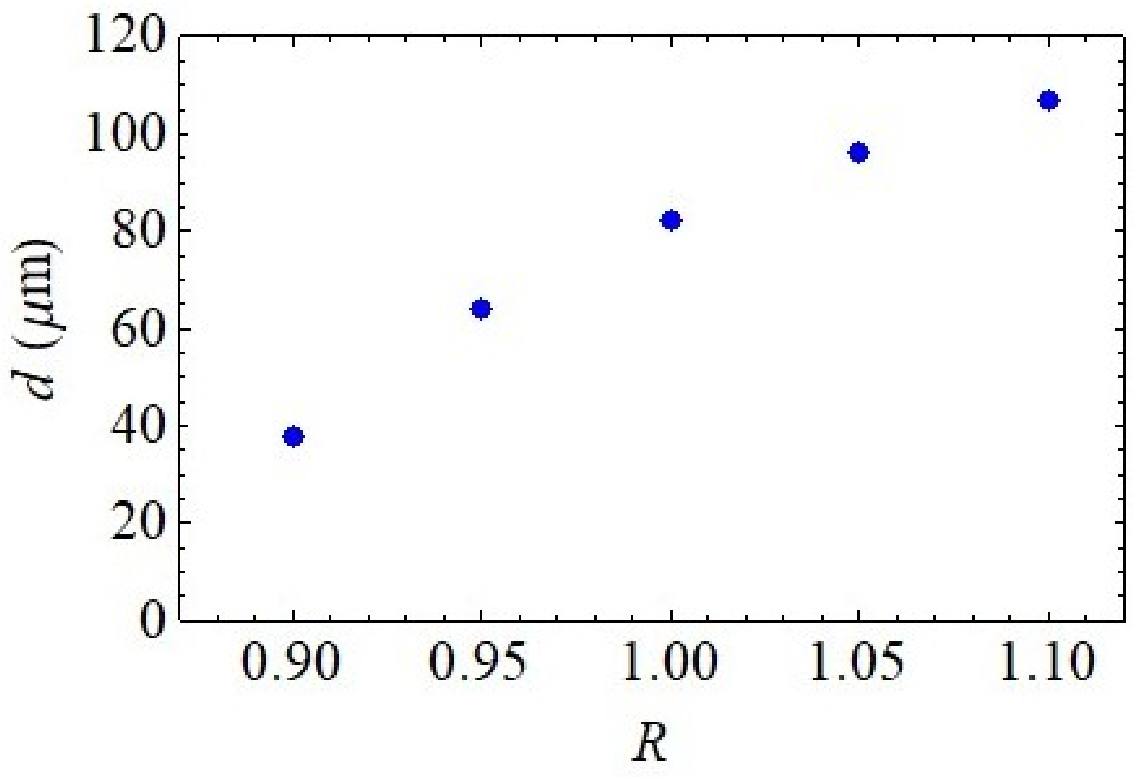}
\caption{Distance between two double-well versus the ratio $R$}
\label{distance} 
\end{figure}

The trap design is based on our linear-type surface-electrode trap~\cite{Tanaka12}. To realize parallel ion strings, it is necessary to produce two RF nodal lines, that is, a double-well pseudopotential in the radial direction. Our trap enables the variation from a single-well to double-well potential by adjusting the applied RF voltages. The design principles of our trap were reported in ~\cite{Tanaka14}. An improved fabrication process, with tighter geometric tolerances and hidden insulator structures, allows us to down-size the ion trap, to reach a regime interesting for the study of many-body physics and nanofriction. We describe the geometry and design of the new ion trap and  focus on the parameters which change the spacing between ion strings. Figure \ref{TrapLayout}(a) shows a schematic illustration of the trap layout. In a usual single-well trap, an RF voltage is applied to the two RF electrodes to provide radial confinement. To produce a double-well potential, an additional RF electrode is located at the center of the trap, which we call center-RF electrode. We apply an RF voltage to the center-RF electrode ($V_{cRF}$) at a certain ratio in addition to the original RF electrodes ($V_{RF}$). The effect of the $V_{cRF}$ is shown in Fig. \ref{TrapLayout} (b)-(d). These calculations are performed for a trap having the same dimensions as the trap developed in this work, but the same behavior of the RF node is observed with traps of different dimension. The widths along the x direction at the trap center of the RF, side, and center-RF electrodes are set to be \SI{409}{\micro\meter}, \SI{26}{\micro\meter}, \SI{78}{\micro\meter}, respectively. We assume to apply RF voltage with the amplitude of 85 V at 27.2 MHz. The ratio $R=V_{cRF}/V_{RF}$ changes the potential. When $V_{cRF}$ is equal to zero, it functions as a usual linear trap, that is, a single RF nodal line is located above the trap surface on the y-axis (Fig. \ref{TrapLayout} (b)). The field from the center-RF electrode is in the positive direction on the y-axis, which reduces the magnitude of the field below the original RF nodal line and increases it above it. The additional $V_{cRF}$ thus makes the RF nodal line lower, but it remains on the y-axis while the field from the outer RF electrode is dominant. As the $V_{cRF}/V_{RF}$ ratio increases, the field from the centre-RF to the next electrode becomes effective and the two parallel nodal lines form near the trap surface. Thus, as the ratio $R$ is increased, the second psudopotential extremum appears along the y-axis first (Fig. \ref{TrapLayout} (c)). Further increasing the ratio changes the configuration drastically. The two nodal lines lay in a plane parallel to the trap surface (Fig. \ref{TrapLayout} (d)).  
 In this case, the distance between two nodal lines $d$ can be adjusted by changing $R$. 
 Figure \ref{distance} shows the calculated distance between two nodal lines $d$ versus $R$.  The trap dimension and RF voltage are same for the case in Fig. \ref{TrapLayout}. When $R$ is set to be 0.9, the distance $d$ and the potential barrier are estimated to be \SI{38}{\micro\meter} and 0.6 meV, respectively. In terms of temperature, this potential barrier corresponds to about 7 K, which is high enough for ions after Doppler cooling.

\begin{figure}
\centering
\includegraphics[width=0.5\columnwidth]{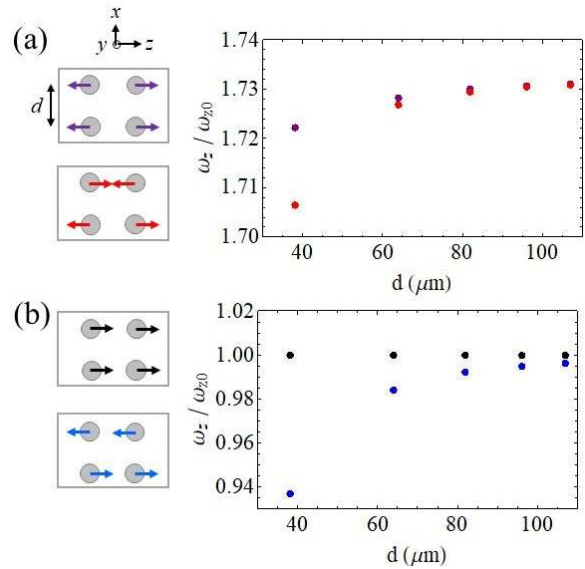}
\caption{Ion motion and corresponding eigenfrequencies of (a) stretch modes and (b) COM modes in the z direction of 2$\times$2 $^{40}$Ca$^+$ ion crystals. The distance between two ion strings is indicated by $d$. A confinement with $\omega_{z0}/2\pi$ of 0.19 MHz is assumed for the calculation. The eigenfrequencies $\omega_{z}$ are normalized by $\omega_{z0}$.}
\label{eigenmodes} 
\end{figure}

The nanofriction emulation using two ion strings is based on Coulomb interaction between these strings. Measurements of vibrational mode frequencies are one of the methods of observing the interaction. We calculated the secular frequencies of 2$\times$2 ion crystals of $^{40}$Ca$^+$ ions at various distances $d$ between the ion strings~\cite{Wunderlich11}. In each direction, there are four eigenmodes. The ions in one string have a center-of-mass (COM) mode and a stretch mode, and each mode has in-phase and out-of-phase configurations between the two strings. Figure \ref{eigenmodes} shows the motion of the eigenmodes and the corresponding calculated eigenfrequencies versus $d$ in the z direction. Because the double-well potential varies when changing $R$, we calculated $\omega_x$ and $\omega_y$ for each $R$. The axial frequency $\omega_{z0}/2\pi$ is set to be 0.19 MHz. The eigenfrequencies $\omega_{z}$ are normalized by $\omega_{z0}$, therefore the value of the in-phase, COM mode corresponds to 1. The distance between two ions in the same string is calculated to be  \SI{17}{\micro\meter}. When $d$ is large enough that the interaction between two ion strings is negligible, both in-phase and out-of-phase modes are degenerate because the two strings are almost independent. The lift of the degeneracy is observed when $d$ is about \SI{40}{\micro\meter}, which is achievable with the trap fabricated in this work.

\subsection{Trap Fabrication}

\begin{figure}
\centering
\includegraphics[width=1\columnwidth]{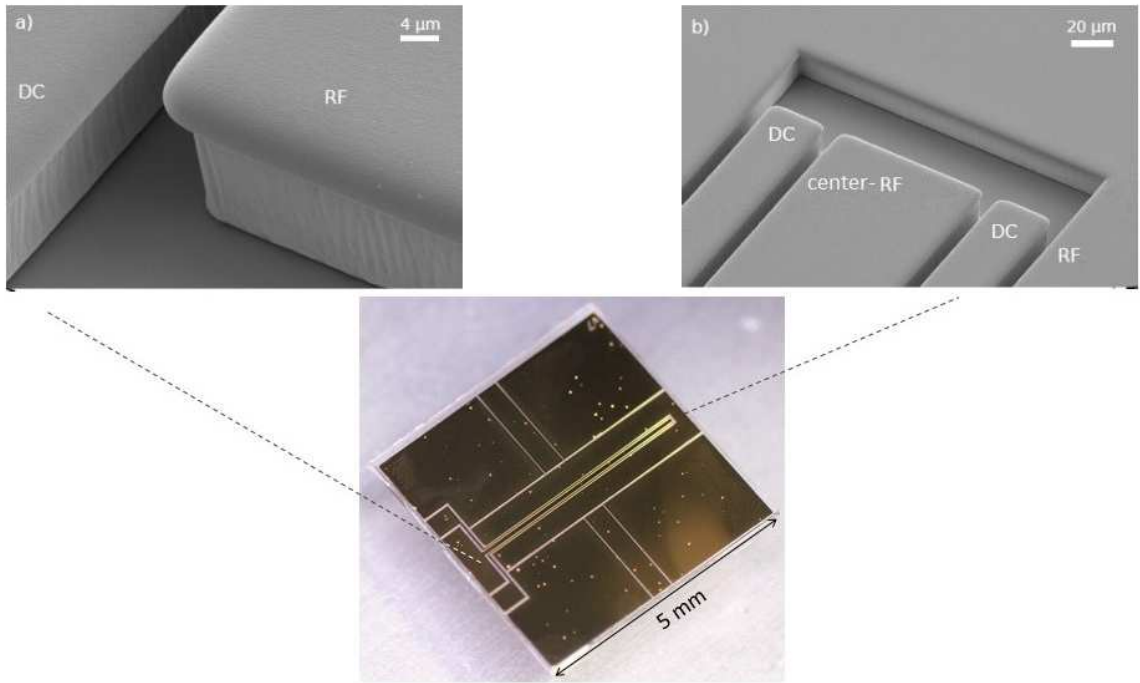}
\caption{A 5~x~5~mm$^2$ trap chip made of gold structures with high aspect ratio (5:1) electroplated on sapphire. a) SEM micrograph of a control DC electrode and an RF electrode separated by an effective gap of \SI{4}{\micro\meter}. b) SEM micrograph of an outermost part of the chip showing part of the RF and DC trap electrodes.}
\label{Thick_Electrodes_On_Sa} 
\end{figure}

In a 3 inch fabrication line located at Physikalisch-Technische Bundesanstalt (PTB) we use a Single Level Processing (SLP) method to produce the trap presented in this work. A sapphire wafer with a thickness of \SI{430}{\micro\meter} or an AlN wafer with a thickness of \SI{635}{\micro\meter} is coated with a thin layer (\SI{12}{\nano\meter}) of evaporated titanium to ensure good adhesion between subsequent gold layers and the substrate. Immediately afterwards a gold layer (\SI{50}{\nano\meter}) is deposited evaporatively to act as a seed layer during the subsequent electroplating process. To define the trap structure we use a photolithography step. At this stage the wafer is spin coated with a positive photoresist (thickness of \SI{16}{\micro\meter}) and then exposed to UV light using contact lithography. Once the resist layer has been developed, we place the patterned wafer in a sulphite-based gold bath and electrochemically deposit gold with a thickness of up to \SI{20}{\micro\meter}. The electroplating process is stopped after a desired overgrowth is reached. After that we remove the remaining developed resist with aceton and isopropanol. As a final step, we etch resist residues under an oxygen-based plasma and remove both gold seed and titanium adhesion layers by means of Ar reactive ion etching. For the case of a positive photoresist, our SLP method allows the fabrication of large (height-to-width) aspect ratio structures. We also minimize the surface roughness after optimizing the applied current densities during electroplating. An example of a high aspect ratio and optimized surface quality structure is depicted in an SEM micrograph shown in Fig.~\ref{Thick_Electrodes_On_Sa}. One important feature is that trap electrodes are as thick as \SI{20}{\micro\m} with gaps of about \SI{4}{\micro\m}, thereby virtually eliminating patch potential from the insulator gaps and thus demonstrating the capability to build gap structures with high aspect ratio of up to 5:1.

\section{Ion Trapping}
 
The fundamental characteristics of the fabricated trap are evaluated with $^{40}$Ca$^+$ ions. The trap chip is mounted in a ceramic pin grid array and set in a vacuum chamber at the vacuum level of 7$\times$10$^{-8}$ Pa. The RF voltages are generated with a dual channel function generator followed by amplifiers. The phase between the two outputs is adjustable with the dual channel function generator. Both amplifier outputs are led to an electric circuit consisting of two transformers, one of which is connected to a circuit for additional DC voltage to the center-RF electrode. At the output to the RF electrodes, a variable capacitor is connected so that the resonance frequency can be tuned to that of the other circuit for the center-RF electrode.
Axial confinement is achieved by applying DC voltages to the end electrodes. The side electrodes located between the RF electrodes and the center-RF electrode are used to apply voltages so that DC potential extrema match the two RF nodal lines~\cite{Tanaka14}. A different axial confinement for each RF nodal line is possible by adjusting the voltages of end electrodes. 
An oven containing calcium atoms is directed toward the trap center along the $z$ axis. Calcium atoms are photoionized with 423-nm and 375-nm lasers. The $^2S_{1/2}-^2P_{1/2}$ transition at 397 nm is used for laser cooling and detection of $^{40}$Ca$^+$ while the $^2D_{3/2}-^2P_{1/2}$ transition at 866 nm is used for pumping back ions from the $^2D_{3/2}$ state to the $^2S_{1/2}-^2P_{1/2}$ transition. The 397-nm laser beam is divided into two beams and one is overlapped with other laser beams. These beams are irradiated along the $x$ axis. The other 397-nm beam propagates at an angle of 45 to the $z$ axis. The fluorescence of the ions is collimated and detected with an EMCCD camera.

\begin{figure}
\centering
\includegraphics[width=0.25\columnwidth]{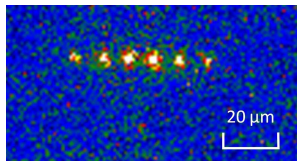}
\caption{Ion string of six calcium ions in the case of single-well operation.}
\label{ionstring} 
\end{figure}

\begin{figure}
\centering
\includegraphics[width=0.8\columnwidth]{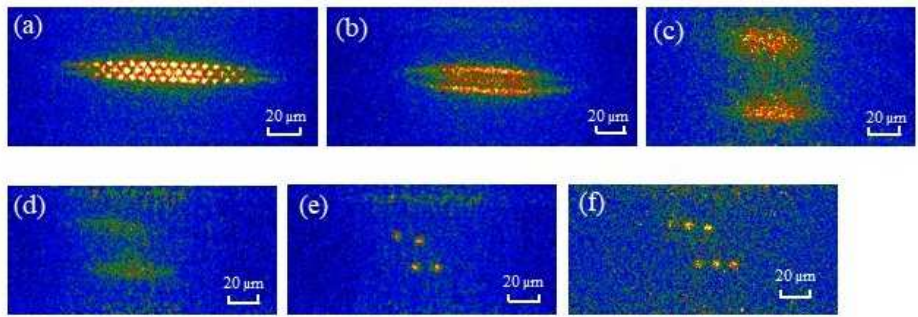}
\caption{Images of ions obtained with $R$ values of (a) 0.60, (b)0.83, (c)0.93, (d)0.85, (e)0.85 and (f)0.88. See text about the error for the $R$.}
\label{doublewell} 
\end{figure}

Figure \ref{ionstring} shows an image of an ion string when the trap is operated in single-well condition at the ratio $R$= 0.07. The frequency and amplitude of the RF voltage were set to be 27.2 MHz and 78 V, respectively. Figure \ref{doublewell} shows ion images observed with various values of $R$. With $R$ as low as 0.60, ions are trapped around the center of the electrode (Fig. \ref{doublewell}(a)). As $R$ is increased up to 0.83, the center part of the ion Coulomb crystals disappears (Fig. \ref{doublewell}(b)), then splits into two parts. Figure \ref{doublewell}(c) shows an image when $R$ is set to be 0.93, which represents the radial potential forming a double-well potential. The states from Figs. \ref{doublewell}(a) to (c) are reversible. At $R=0.85$, the distance between two wells becomes closer as shown in Fig. \ref{doublewell}(d). After reduction of the ion number by blocking the cooling laser beam several times, we observe two-dimensional ion crystals (Fig. \ref{doublewell}(e)). Figure \ref{doublewell}(f) shows parallel ion crystals obtained in the same manner as Fig. \ref{doublewell}(e) for a different $R$.

The observed ion configurations show that the trapping potential changes
from single-well to double-well by changing the ratio $R$ as expected.
However, we found that the relation between the value of $R$ and the ion configuration is in disagreement with the simulation. The value of $R$ at which the two RF nodal lines become to lay in a plane parallel to the trap surface is estimated to be between 0.85 and 0.88 for this electrode dimension, whereas ions begin to split into two parts at 0.83 (Fig. \ref{doublewell}(b)) and form two dimensional configuration even at 0.85 (Fig. \ref{doublewell} (d), (e)). We consider that this is mainly due to an error in voltage measurements. The voltages applied to the RF and center-RF electrodes are measured at a feedthrough outside the vacuum chamber with high-voltage probes. From measurements of the secular frequencies in the case of single-well operation, we infer that the measured RF voltages differ from the actual voltages by at least 3\%, which introduces an error in $R$ of approximately $\pm$0.05 in the worst case.

To achieve crystallization of any number of ions at a certain $d$, fine adjustment of the applied DC voltages is necessary so that the null points of the electric field from the DC potential match the rf potential nodes. We apply DC voltages to four end, two middle, and two side electrodes, and add a DC voltage to the center-RF electrode. Appropriate voltages for the nine DC electrodes are derived by using the method of Lagrange multipliers ~\cite{Tanaka14}. In addition, the effect of a stray electric field must be taken into account and compensation voltages need to be applied. At present, the adjustment of DC voltages has been performed by changing the output of a DA converter manually. We plan to develop an automatic DC voltage adjustment system where a stray electric field around the trapping region is estimated from an ion image then appropriate DC voltages are provided after the calculation of compensation voltages.

\section{Nanofriction emulator}

Parallel ion strings can be regarded as two chains of ions in which the ions couple to each other due to their Coulomb interaction and as such are exposed to a periodic potential produced by the other ion string\cite{Mehlstaubler17}. This situation is similar to the Frenkel-Kontorova model (FK model)  ~\cite{FKmodel} which is one of the successful models in the study of nanofriction. In the FK model, a string of coupled particles is subject to a fixed sinusoidal corrugation potential mimicking the surface of a solid-state body. When the string slides over the corrugation potential, the motion is affected by the equilibrium spacing of the string, and the periodicity and amplitude of the corrugation potential.  In the case of an infinite system of incommensurate periodicities, a transition from sliding to stick-slip motion occurs, which is known as the Aubry transition~\cite{Aubry83}. The behavior of such a system is characterized by the corrugation parameter which is defined as $\eta=(\omega_{int}/\omega_0)^2$~\cite{Vuletic15, Vuletic17, Mehlstaubler17, Schmirander17}, where $\omega_{int} $ is the frequency of vibration of an ion in a potential generated by an adjacent ion string and $\omega_0$ is the frequency of vibration in a potential produced by ions in the same string. To investigate this friction phenomenon, it is interesting to set the corrugation parameter $\eta$ to be around 1, where the interaction with all ions of the same string and that with ions of the other string are comparable, and the transition between free sliding and the stick-slip mode occurs. 

\begin{figure}
\centering
\includegraphics[width=0.6\columnwidth]{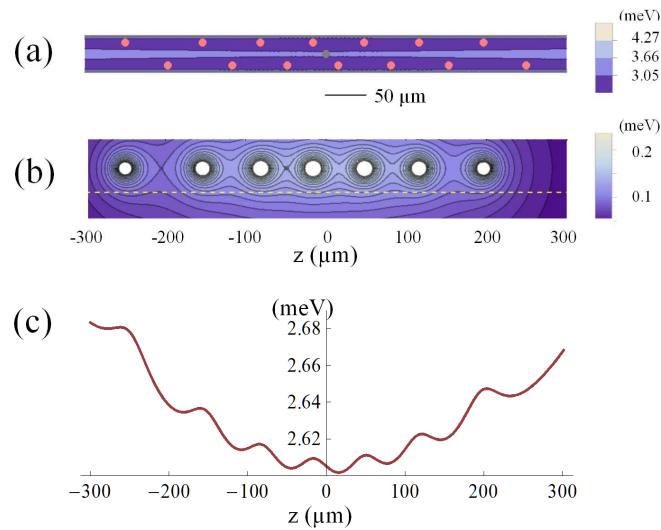}
\caption{Calculated corrugation potential for 14 ions in a double-well trapping potential. (a) Equilibrium positions of 14 ions in a double-well potential (pink dots). The gray dot at the center represents the origin. The frequency and amplitude of RF voltage are set to be 27.2 MHz and 120 V, respectively. The axial trapping frequencies of both potential wells are set to be 15 kHz. The spacing between two ion strings is \SI{30}{\micro\meter}.  (b) Contour plot of the Coulomb potential energy produced by the upper ion string.The yellow dashed line represents the position of the other ion string. (c) Potential energy at the lower ion string whose position is indicated the yellow dashed line in (b). The curve represents the potential including both Coulomb potential produced by the upper ion string and the external trapping potential. }
\label{corrugation} 
\end{figure}

We estimated $\eta$ of several configurations for parallel ion strings. Figure  \ref{corrugation} shows one of the simulated configurations of parallel ion strings consisting of 14 ions, which provides $\eta \simeq 1$. In this simulation, the frequency and amplitude of the RF voltage are set to be 27.2 MHz and 120 V, respectively. The axial trapping frequencies of both potential wells are set to be 15 kHz. The spacing between two ion strings is  \SI{30}{\micro\meter}. The barrier between two rf pseudopotential wells is estimated to be 0.9 meV at this RF voltage. $\omega_{int}$ is obtained from this curve by applying a harmonic approximation to the center well and using ${\omega_{int}}^2=\frac{1}{m} \frac{\partial^2 U}{\partial  z^2}$, where $U$ is the harmonic potential and $m$ is the ion mass. Similarly, $\omega_0$ is obtained from the calculated potential at the center ion of the lower string produced by other ions in the same string. In the case of Fig. \ref{corrugation}, $\omega_{int}/2\pi$ and $\omega_0/2\pi$ are estimated to be 42.1 kHz and 41.9 kHz, respectively.  The corrugation parameter $\eta$ is adjusted by varying the distance $d$ between two strings, that is, by changing the ratio $R$ of the RF voltages. For the above given parameters, $\eta$ can be changed from 0.57 to 1.65. Further modification is also possible by adjusting the axial trapping potential as well as $d$. Loading different numbers of ions into the two RF traps will allow to tune the periodicities of both crystals and alter the frictional dynamics. Both crystals can then be moved against each other by changing the end cap electrodes. The two sliding and interacting crystals will form a self-assembled Coulomb system with backaction. Ultimately, the system can be scaled to two- or three-dimensional crystals with intriguing complex dynamics.
It will be possible to investigate the dynamical behavior of particles in the FK model under various temperature conditions, observe the Aubry-like transition with low-temperature ions, and study the thermolubricity. The independent control of ion crystals in both RF traps allows for versatile experiments and tests of nanofriction, which can be scaled in this setting from low dimensional strings as found in DNA and biomolecules up to 2D and 3D systems interacting with each other. Note, that by choosing a different axial confinement or different ion numbers for the two RF traps, a different dimensionality of crystals can be obtained in the different traps. Like this 1D or 3D objects can slide against 2D surfaces.

\section{Conclusions}
We have discussed an application of a microfabricated ion trap that produces a double-well potential to implement a nanofriction emulator. Our trap design enables the trapping potential to be varied from single-well to double-well configuration in the radial direction by varying an additional RF voltage on the center electrode. In addition, the distance $d$ between the two potential wells is adjusted by the ratio $R=V_{cRF}/V_{RF}$. We have designed a trap half the size of our previous trap ~\cite{Tanaka14} so that the interaction between two ion strings becomes significant. With this trap, the degeneracy of the eigenmode frequencies for 2$\times$2 ion crystals of $^{40}$Ca$^+$ ions is lifted when the distance $d$ is around \SI{40}{\micro\meter}. We have reported the process to fabricate such a downsized trap and achieved an aspect ratio of electrode thickness to the gap as high as 5:1. With this microfabricated trap, calcium ions have been trapped under both single-well and double-well conditions. As $R$ is increased, we have observed that ions are split into two parts. Small number of ions have been crystallized, however, further fine adjustment of the applied DC voltages is required for the formation and manipulation of large number ion crystals. We plan to develop a system in which the electric field in the trapping region is derived from an ion image, then compensation voltages are calculated and applied to the DC electrodes. Finally, we have shown that our system can access the corrugation parameter $\eta \simeq 1$. The trap developed in this work has a large enough tuning range in $\eta$, which makes it a promising system for the study of nanofriction with laser cooled and trapped ions. 
We propose using such a platform as a versatile tool where the periodicity of the  ion strings can be tuned independently of each other and crystals moved against each other. Another degree of freedom comes in, when different numbers of ions are loaded directly into the RF trap minima, e.g.  by two focused photo-ionization laser beams with fast shutters.
Last but no least, the system of study can be low-dimensional strings as found in DNA and biomolecules or, depending on the aspect ratio of axial and radial trapping frequencies, 2D and 3D systems interacting with each other. Ultimately, 1D or 3D objects can slide against 2D surfaces and their dynamics be studied spectroscopically with atomic resolution.
\\
\ack
This work was carried out within the 
 international joint research lab  on trapped-ion integrated atomic-photonic circuits $``$TRiAC", stimulated by the collaborative grant JSPS-DAAD and Osaka University's International Joint Research Promotion Program. This work was supported by JST CREST under grant number JPMJCR1776, JAPAN and through Germany's Excellence Strategy  EXC-2123 QuantumFrontiers - 390837967.
\\

\textbf{References}

\bibliography{references}

\end{document}